\RequirePackage{fix-cm}
\documentclass[
 aps,prl,
 amsmath,amssymb,superscriptaddress,
 preprint,
]{revtex4-1}

\usepackage{graphicx}
\usepackage{dcolumn}
\usepackage{bm}
\usepackage{color}

\usepackage[utf8]{inputenc}
\usepackage[T1]{fontenc}
\usepackage{mathptmx}

\usepackage[normalem]{ulem}
\begin{document}

\title{Onset of metallic transition in molecular liquid hydrogen}

\author{Jianqing Guo}
\affiliation{School of Physics, Peking University, Beijing 100871, People’s Republic of China}
\affiliation{International Center for Quantum Materials, Peking University, Beijing 100871, People’s Republic of China}%

\author{Bingqing Cheng}
\affiliation{The Institute of Science and Technology Austria, Am Campus 1, 3400 Klosterneuburg, Austria}

\author{Limei Xu}
\affiliation{School of Physics, Peking University, Beijing 100871, People’s Republic of China}
\affiliation{International Center for Quantum Materials, Peking University, Beijing 100871, People’s Republic of China}%
\affiliation{Interdisciplinary Institute of Light-Element Quantum Materials and Research
Center for Light-Element Advanced Materials, Peking University, Beijing, China.}

\author{Enge Wang}
\affiliation{School of Physics, Peking University, Beijing 100871, People’s Republic of China}
\affiliation{International Center for Quantum Materials, Peking University, Beijing 100871, People’s Republic of China}
\affiliation{Interdisciplinary Institute of Light-Element Quantum Materials and Research
Center for Light-Element Advanced Materials, Peking University, Beijing, China.}
\affiliation{Songshan Lake Materials Lab, Institute of Physics, Chinese Academy of Sciences, Guangdong, China.}
\affiliation{
School of Physics, Liaoning University, Shenyang, China
}

\author{Ji Chen}
\email{ji.chen@pku.edu.cn}
\affiliation{School of Physics, Peking University, Beijing 100871, People’s Republic of China}
\affiliation{Interdisciplinary Institute of Light-Element Quantum Materials and Research
Center for Light-Element Advanced Materials, Peking University, Beijing, China.}
\affiliation{
 Collaborative Innovation Center of Quantum Matter, Beijing 100871, People’s Republic of China
}

\date{\today}

\begin{abstract}
Liquid-liquid phase transition of hydrogen is at the center of hydrogen phase diagram as a promising route towards emergent properties such as the Wigner-Huntington metallization, superconductivity, and superfluidity. 
Here we report a study on the liquid-liquid phase transition of hydrogen using the state-of-the-art diffusion quantum Monte Carlo and density functional theory calculations.
Our results suggest that the metallization process happens at lower pressures and temperatures compared to the structural phase transition of molecular to atomic hydrogen.
The consequence is that metallized molecular hydrogen is stable at a wide range of pressures and temperatures.
Our study breaks the conventional assumption that metallization coinciding with dissociation of hydrogen molecule, and the molecular metallic hydrogen liquid phase is likely to become the frontier of studying hydrogen phase transitions.

\end{abstract}

\maketitle

Hydrogen, the simplest and lightest element in nature, is famous for a complex phase diagram and a rich variety of phase transitions.
Among the many phases of hydrogen, the liquid-liquid phase transition (LLPT) is of particular interest ever since metallic hydrogen was predicted in 1935, and LLPT is considered as a path to metallic hydrogen \cite{mcmahon_properties_2012,nellis_perspective_2021}.
Experimentally, the LLPT of hydrogen has been achieved under dynamic and static compression in the fluid phase, and the evidence of metallization was mostly provided by the change in optical reflectivity  \cite{dzyabura-pnas-2013,zaghoo-prb-2016,zaghoo-prb-2018,jiang-advsci-2020,weir-prl-1996,knudson-science-2015,peter-science-2018}.
Nevertheless, challenges and controversies still exist in experiments especially at low temperatures, including the uncertainty in determining the extreme pressure and temperature conditions, the difficulties in structure characterizations, and the lack of accurate transport measurements of conductivity \cite{mcmahon_properties_2012,fang_quantum_2019,silvera_phases_2021}. 

Theoretically, 
LLPT of hydrogen is often described as a transition from the insulating molecular phase to the metallic atomic phase, supported by computational simulations \cite{scandolo-pnas-2003,holst-prb-2008,morales_evidence_2010, lorenzen-prb-2010,morales-prl-2013, geng-prb-2019,hinz-prr-2020,tian-jpcm-2020,vandebund-prl-2021}, indicating or assuming that the molecular-atomic transition (MAT) and the insulating-metallic transition (IMT) occur simultaneously.
In recent years, new attention has been paid to the supercritical behavior and the location of liquid-liquid critical point in the computational community, involving studies using density functional theory (DFT), machine learning potentials and quantum Monte Carlo (QMC) \cite{li_supercritical_2015, gorelov-prb-2020, cheng-nature-2020, karasiev_liquidliquid_2021,Cheng2021_reply, tirelli_high_2021}. 
Based on these studies, it is understood that 
below the critical point, if the temperature is above the melting line,
LLPT is one phase transition where $\text{H}_2$ molecule dissociates and metallizes simultaneously; but above the critical point, the dissociation and metallization happen in a smooth and continuous manner.
In literature, both the molecular-atomic transition and the insulating-metallic transition have been analysed in detail using the state-of-the-art computational methods (see e.g. Refs. \onlinecite{pierleoni_local_2018, gorelov-prb-2020}).
However, to the best of our knowledge, there is no theoretical guarantee or experimental evidence that the metallization of molecular hydrogen should exactly coincide with the dissociation, regardless if LLPT is first order or not.
On the contrary, experimental studies performed in the solid regime of dense hydrogen has suggested that metallization can happen without a molecular-atomic transition \cite{eremets_semimetallic_2019}.
{Furthermore, it is worth mentioning that Mazzola et al. predicted a mixed molecular-atomic(MA) liquid phase using QMC-MD simulations based on an efficient generalized second-order Langevin dynamics \cite{mazzola-prl-2015}. The existence of MA phase can directly prove that the metallization at finite temperature occurs before the complete dissociation of the liquid hydrogen. 
However, due to the possibility of errors in their previous simulations, the predicted metallization pressure was abnormally high and the existence of the MA phase was then excluded in their later work \cite{mazzola_prl_2018}. These intriguing experimental and theoretical results need to be further studied, motivating a re-examination of the nature of LLPT.}

In this paper, we study the molecular-atomic and insulating-metallic phase transition of liquid hydrogen, to answer the basic question of whether they coincide with each other or one occurs before another. 
To provide microscopic insights, we propose the molecular fraction of liquid hydrogen as the key order parameter to compare the two kinds of phase transitions on the same foot. 
We use the state-of-the-art diffusion quantum Monte Carlo (DMC) method to benchmark DFT functionals and find that vdw-DF can accurately describe the structural phase transition of dense liquid hydrogen. 
We also study the electronic phase transition of liquid hydrogen by calculating the fundamental gap as a function of molecular fraction using DMC and DFT methods. 
We find that the insulating-metallic transition begins when around 20\% hydrogen molecules dissociate into hydrogen atoms, which requires much lower dissociation than the half molecular-atomic phase transition. 
By mapping such results onto the pressure-temperature phase diagram, we estimate that the pressure of IMT is tens of GPa lower than structural MAT.

\begin{figure*}
\includegraphics[width=6.69 in]{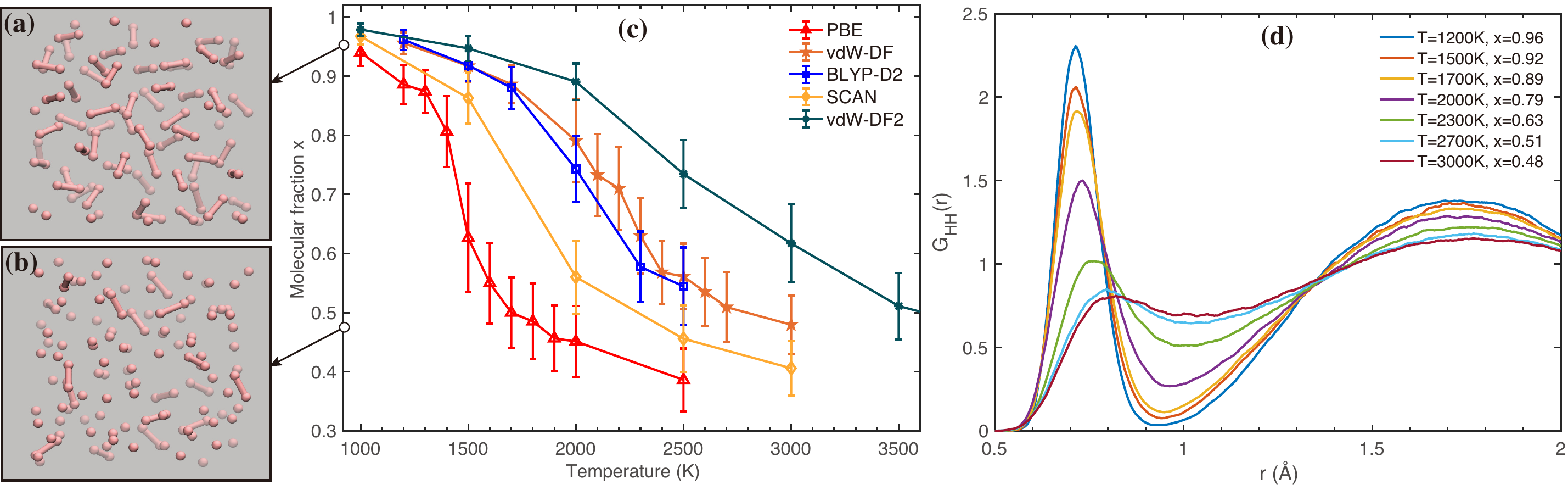}
\caption{
 Molecular and atomic liquid hydrogen from AIMD simulations.
 (a-b) Snapshots of typical molecular liquid hydrogen (x = 0.96) and half-molecular-half-atomic liquid hydrogen (x = 0.48) from the vdW-DF simulations. The criterion for drawing the H-H bond is 0.9~\AA.
 (c) Molecular fraction as a function of temperature using different DFT functionals. 
 The molecular fraction order parameter x is defined in detail in the main text and the Supplementary Material. Qualitatively the molecular faction x describes the portion of $\text{H}_2$ molecules in liquid hydrogen.
 (d) HH pair correlation functions $\text{G}_\text{HH}(\text{r})$ at different temperatures calculated from vdw-DF AIMD simulations.
 Simulations were carried out at the density of $\text{r}_\text{s}$ = 1.50 Bohr in the \textit{NVT} ensemble with 128 hydrogen atoms. 
 }
\label{fig:1} 
\end{figure*}

DFT-based \textit{ab initio} molecular dynamics (AIMD) simulations were carried out using the Quantum Espresso (QE) package \cite{giannozzi-jpcm-2009,giannozzi-jpcm-2017} with the Perdew-Burke-Ernzerhof (PBE) \cite{perdew-prl-1996}, BLYP-D2 \cite{grimme_blypd2_2006}, the vdW-DF \cite{dion_vdwdf_2004}, the vdW-DF2 \cite{klime_vdwdf2_2009} and the SCAN \cite{sun-prl-2015} functionals.
The interactions between the valence electrons were treated with Hamann-Schl\"uter-Chiang-Vanderbilt pseudopotentials \cite{hamann-prl-1979,vanderbilt-prb-1985}. 
Each simulation was performed in the canonical (\textit{NVT}) ensemble, and the temperature was controlled by a Stochastic Velocity Rescaling thermostat \cite{bussi2007svr}.
DFT total energy calculations with the PBE, the vdW-DF, the BLYP-D2, the SCAN, the HSE06 \cite{heyd_hse06_jcp} functionals were performed with both the QE package and the VASP code \cite{kresse_vasp_1996}.
DMC calculations were performed using the CASINO package \cite{needs_variational_2020}, with the size-consistent DMC algorithm ZSGMA \cite{zen-prb-2016}.
The recently developed energy consistent correlated electron pseudopotentials (eCEPPs) \cite{trail-jcp-2017} were used.
Further computational details can be found in the Supplemental Material \cite{si}.

To investigate the molecular-atomic phase transition we use an order parameter, namely the molecular fraction.
%
As suggested in previous studies, a molecular fraction of 0.5 coincides with the pressure plateau, which is a common criterion to identify the molecular-atomic phase transition \cite{cheng-nature-2020}.
Specifically, we follow the definition of Cheng et al.\cite{cheng-nature-2020}, where the molecular fraction x was defined as the fraction of atoms with one neighbour within a smooth cutoff function starts from 0.8 \AA{} and decays to 0 at 1.1 \AA~(see Supplementary Material II for more details and further validation using an agnostic classification scheme \cite{si}).
There are other definitions of molecular fraction \cite{tamblyn_prl_2010,geng-prb-2019} and other order parameters, e.g. density/volume, heat capacity, compressibility, and diffusion coefficient, that can be used to describe the molecular-atomic transition on the temperature pressure phase diagram \cite{li_supercritical_2015,cheng-nature-2020,karasiev_liquidliquid_2021}.
A detailed discussion on the thermodynamic properties involved in the molecular-atomic LLPT is beyond the scope of this study.
In this study we focus on the molecular fraction, which provides the clearest microscopic insight to the molecular-atomic transition.
Fig. \ref{fig:1} (a) and (b) show snapshot of two structures with a high molecular fraction of (x=0.96) and a low molecular fraction (x=0.48), respectively.

In Fig. \ref{fig:1} (c) we plot the molecular fraction as a function of temperature by performing systematic AIMD simulations using different DFT exchange correlation functionals at the density of $\text{r}_\text{s}$ = 1.50 Bohr with a system size of 128 H atoms.
Tests on e.g. size effects and density dependence are further presented and discussed in Fig. S2 and S3 \cite{si}.
The simulations confirm that the molecular fraction can be used to monitor the molecular-atomic LLPT as a function of temperature.
The structural transition is further illustrated by the pair correlation functions in Fig. \ref{fig:1} (d) and Fig. S1 \cite{si}, where the first peak is suppressed when the molecular fraction decreases to $\sim$0.5. 
However, different functionals show quite different transition temperatures at a constant density, which are consistent with shifts of phase boundaries on a pressure-temperature phase diagram observed in previous studies.  
Among the various functionals considered, we find that PBE predict the lowest phase transition temperature and vdW-DF2 has the highest phase transition temperature.
The other three functionals, namely SCAN, vdW-DF, and BLYP-D2, show behaviors that are in between the cases of PBE and vdW-DF2.
Overall, these computational observations of the qualitative difference among different DFT functionals agree with the comparisons reported in literature \cite{clay_prb_2014, geng-prb-2019}. 
\begin{figure}[t]
 \includegraphics[width=3.37 in]{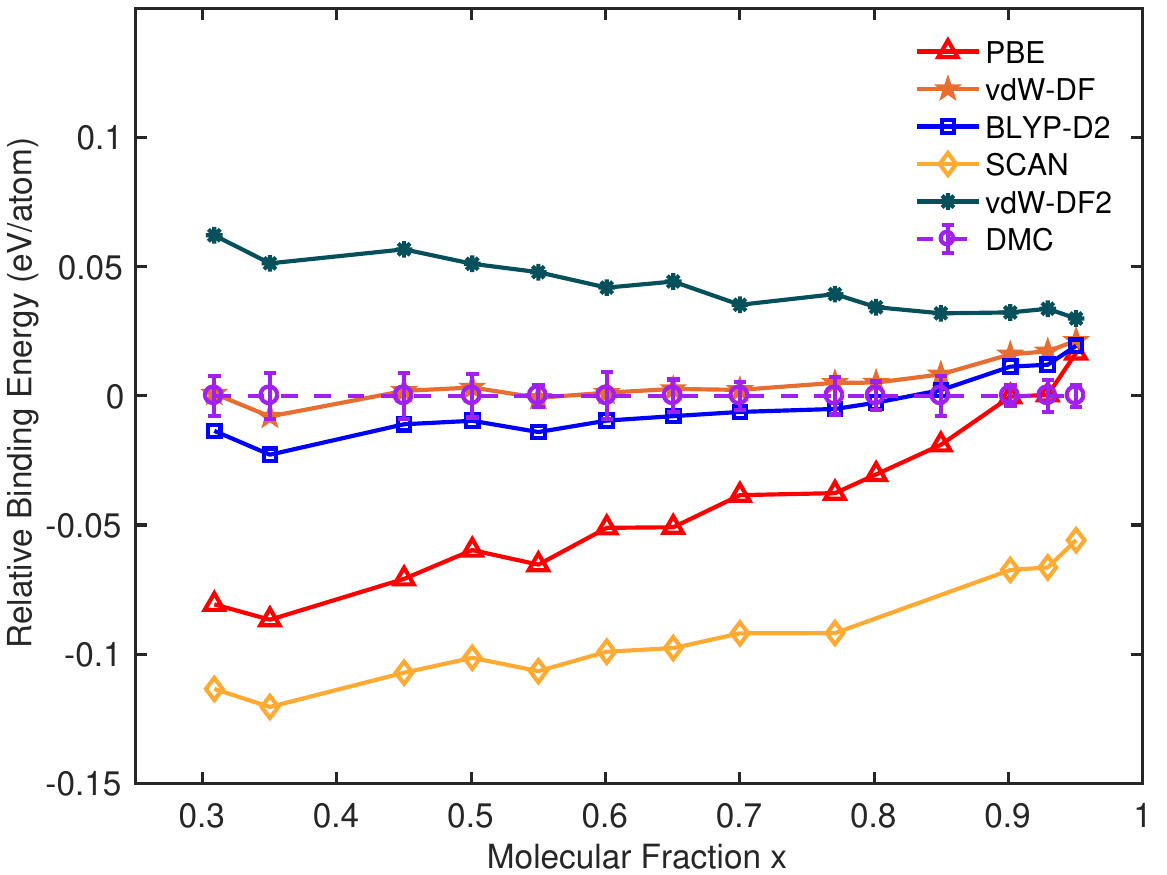}
 \caption{Benchmark of DFT against DMC. 14 structures covering a wide range of molecular fraction x were selected for the benchmark. The structures are collected from vdW-DF AIMD and machine learning potential simulations of 128 hydrogen atoms with a density of rs=1.5 Bohr at different temperatures (see Supplementary Material II for more details \cite{si}).
The relative binding energy is calculated as $\text{E}_\text{b}^{\text{DFT}}-\text{E}_\text{b}^{\text{QMC}}$, where $\text{E}_\text{b}=(\text{E}_\text{liquid} - \text{N}_\text{H} \times \text{E}_\text{H})/\text{N}_\text{H} $ is the binding energy.}
 \label{fig:2} 
 \end{figure} 

The different results from AIMD simulations suggest that it is necessary to further evaluate the accuracy of DFT functionals for liquid hydrogen.
Therefore, we calculate the binding energies of a set of structures via DMC and use them as the benchmark to evaluate the performance of DFT functionals, a strategy that has been adopted successfully for solid hydrogen \cite{chen_jcp_2014, clay_prb_2014, drummond_quantum_2015}.
In Fig. \ref{fig:2}, we plot the relative binding energies of liquid hydrogen structures as a function of the molecular fraction with respect to the DMC calculations ($\text{E}_\text{b}^{\text{DFT}}-\text{E}_\text{b}^{\text{QMC}}$). 
The binding energy is defined as $\text{E}_\text{b}=(\text{E}_\text{liquid} - \text{N}_\text{H} \times \text{E}_\text{H})/\text{N}_\text{H} $, where $\text{E}_\text{liquid}$ is the total energy computed using DFT or DMC, and $\text{E}_\text{H}$ is the energy of a single H atom.
As one can see, vdw-DF and BLYP+D2 describe the interactions in liquid hydrogen most accurately among all the functionals.
From the atomic side to the molecular side, the relative binding energy differs by less than 10 meV per atom, and the minor slope of the curve also benefits the prediction of phase boundary between the molecular phase and the atomic phase.
We also find that PBE and SCAN tend to overestimate the stability of liquid hydrogen, and
a large positive slope means an severe overestimate of the relative stability of the atomic phase.
In contrast, vdw-DF2 underestimates the stability of liquid hydrogen and the underestimation of the stability of the atomic phase is stronger.  
Based on the benchmark, we conclude that vdw-DF and BLYP+D2 describe the structural phase transition of liquid hydrogen most accurately, whereas PBE and SCAN tend to facilitate the molecular-atomic transition and vdw-DF2 hinders the hydrogen molecules from dissociating into atoms. 
The benchmark conclusions are in line with the AIMD results: the vdw-DF and BLYP+D2 results coincide with each other within a small variation; PBE and SCAN predict lower phase transition temperatures while vdw-DF2 has a higher phase transition temperature.
Note that the DFT methods that we used are different in both the nature of the underlying exchange correlation functionals and the van der Waals correction scheme.
The importance of van der Waals interactions in dense hydrogen have been mentioned in some studies, however further analyses show that the differences in describing the molecular and atomic liquid hydrogen are mainly dominated by the underlying exchange correlation functional while the van der Waals interactions do not play a significant role (Fig. S4 \cite{si}). 
\begin{figure}[t]
 \includegraphics[width=3.37 in]{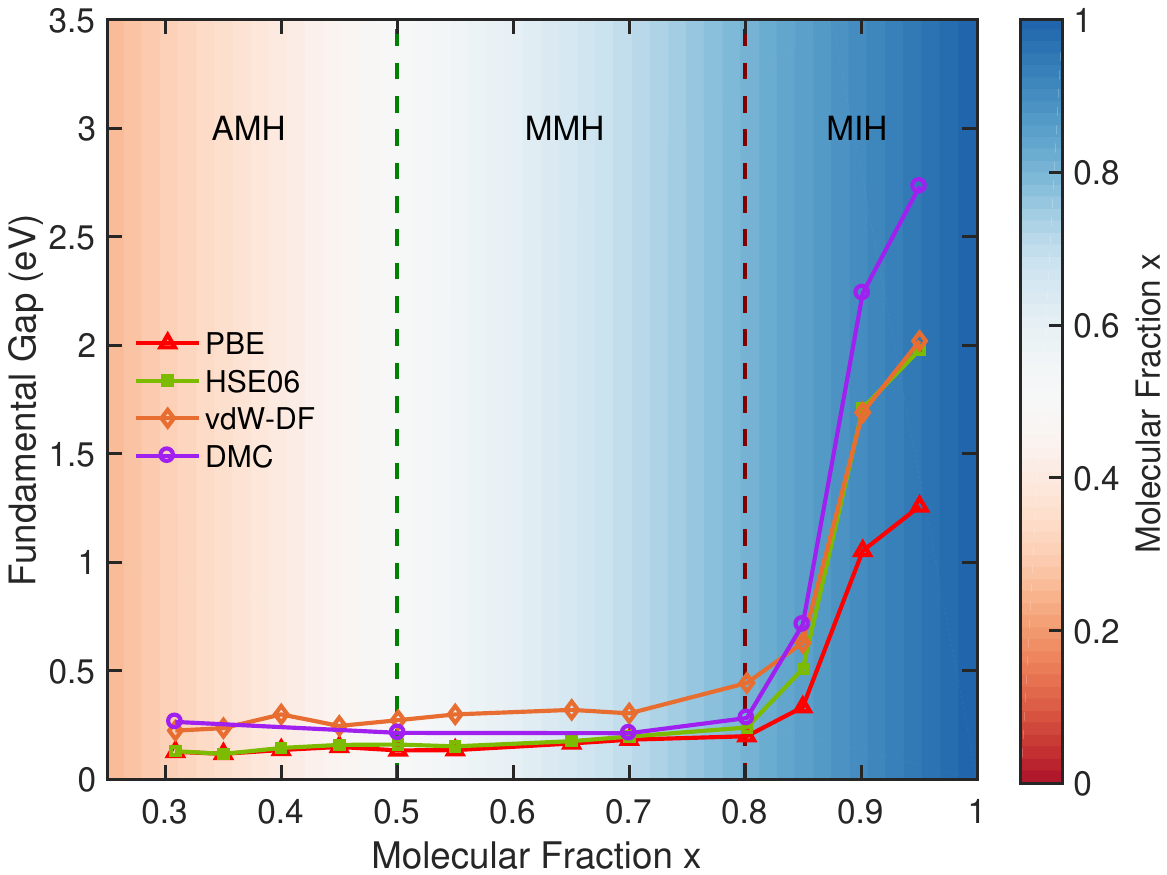}
   \caption{Fundamental energy gap for liquid hydrogen as a function of molecular fraction x calculated using different methods. Fundamental gap is defined as $\Delta = \text{E}(\text{N}+1)+\text{E}(\text{N}-1)-2\text{E}(\text{N})$, where $\text{N}=128$ is the number of electrons, which equals the number of hydrogen atoms in the system. Lines with different color show fundamental gap computed using different methods. The structures are selected from those used in Fig. \ref{fig:2}, and are detailed in the Supplementary Material. AMH, MMH and MIH indicates regimes of atomic metallic hydrogen, molecular metallic hydrogen and molecular insulating hydrogen, respectively. The background is colored with the molecular fraction.}
 \label{fig:3}
\end{figure}

We now consider the electronic insulating-metallic phase transition of liquid hydrogen. 
The fundamental gap of liquid hydrogen as a function of molecular fraction calculated by DMC is shown in Fig. \ref{fig:3}, computed according to $\Delta = \text{E}(\text{N}+1)+\text{E}(\text{N}-1)-2\text{E}(\text{N})$. 
We find that the fundamental gap $\Delta$ closes at molecular fraction $x = 0.8$ (red dashed line), which is far from the half molecular-atomic phase transition boundary $x = 0.5$ (green dashed line). 
As mentioned before, $x = 0.5$ may not be a strict criterion for MAT, but it often coincides well with MAT and can at least be used as a lower bound for MAT, which we will also show in other calculations below.
A fraction of $x = 0.8$ means liquid hydrogen becomes conductive when about 20\% hydrogen molecules dissociate into atomic hydrogen.
Note that the structures used in computing the fundamental gap are collected from simulations covering a wide range of temperatures and pressures \cite{si}. 
Therefore, despite that hydrogen metallizes easier under higher pressures, the molecular fraction can be a robust order parameter to characterize the metallization of hydrogen.
We can also compute the fundamental gap using DFT methods.
As show in Fig. \ref{fig:3}, DFT calculations generally underestimate the fundamental gap compared with DMC.
Specifically, HSE06 and vdW-DF underestimates the gap by up to 30\%, whereas PBE may underestimate by more than 50\%.
However, an interesting fact is that the molecular fraction where the fundamental gap closes is independent of the calculation methods. 
All DFT functionals predict consistently a threshold of molecular fraction $x = 0.8$ for gap closing, therefore we can directly calculate the fundamental gap using more economic methods to test the size effects in fundamental gap calculations. 
So we select more configurations from vdw-DF AIMD simulations with 512 hydrogen atoms at different densities ($\text{r}_\text{s}$ = 1.35 Bohr,$\text{r}_\text{s}$ = 1.50 Bohr and $\text{r}_\text{s}$ = 1.65 Bohr respectively) and calculate their fundamental gaps using PBE, as shown in Fig. S5 \cite{si}. 
We find that the insulating-metallic phase transition boundary is always around molecular fraction $x = 0.8$.
A clear prediction from the above results is that dense liquid hydrogen can be metallized before the molecular-atomic phase transition, which implies that experiments can find metallic hydrogen at lower temperatures/pressures than atomic hydrogen.

\begin{figure*}
\includegraphics[width=6.69 in]{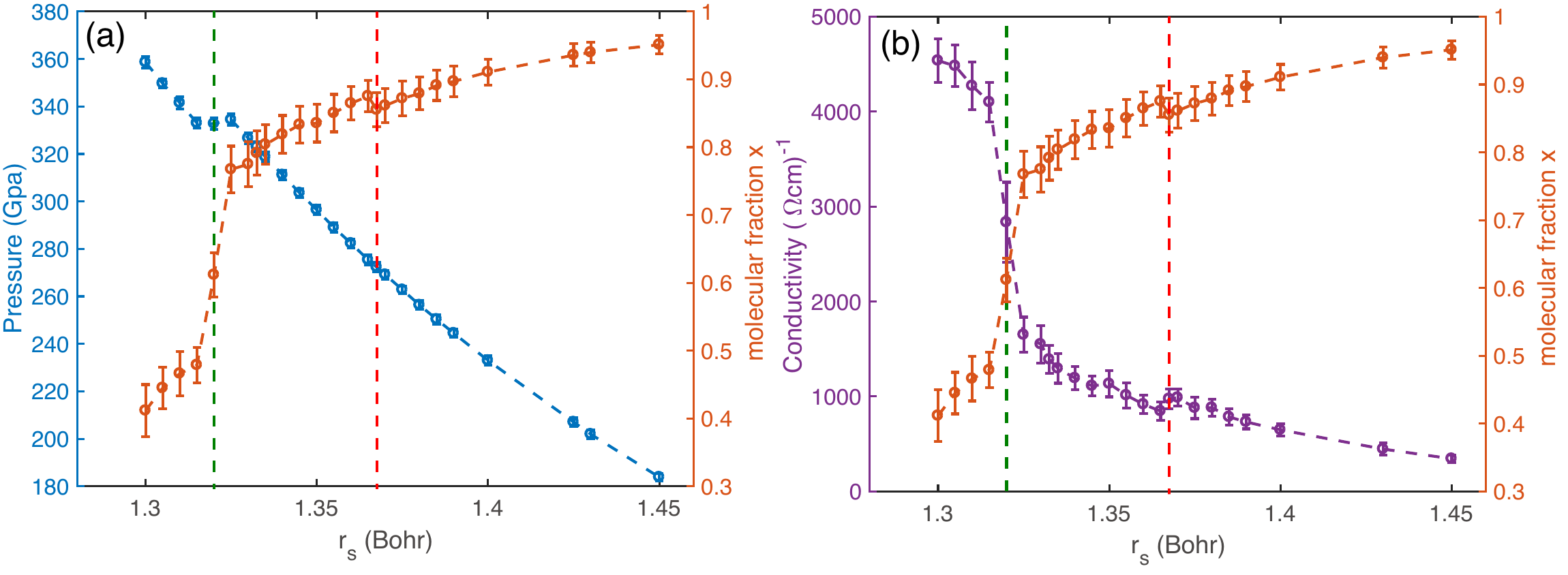}
\caption{
 {Pressure and conductivity of liquid hydrogen from AIMD simulations. All the simulations were carried out at the density of $\text{r}_\text{s}$ = 1.30-1.45 Bohr in the \textit{NVT} ensemble with 256 hydrogen atoms at the temperature of 1000 K.
 The calculated pressure (a) and conductivity (b) as a function of density. The corresponding molecular fractions are also plotted. The vertical green and red dashed lines represents the sudden change of pressure and conductivity respectively.}
 }
\label{fig:eos} 
\end{figure*}

{Furthermore, the insulating-metallic phase transition can be investigated directly by monitoring the electric conductivity of liquid hydrogen. We calculated the pressure and conductivity of liquid hydrogen as a function of density, as shown in Fig. \ref{fig:eos}. Two separate anomalies are clearly found in the conductivity curve while only one transition is found in the pressure. As we already know, the sudden change of pressure represents a transition of atomic structure and the anomaly in conductivity indicates a transition of electronic structure of liquid hydrogen. Thus, our results show that the transition in electronic structure of liquid hydrogen can occur before the atomic structural transition. To be more specific, at the density of $\text{r}_\text{s}$ = 1.37 Bohr with the molecular fraction x = 0.8-0.85, the electric conductivity of liquid hydrogen goes through a sudden change (Fig. \ref{fig:eos} (b)) while the pressure has no abnormal behavior (Fig. \ref{fig:eos} (a)). At the density of $\text{r}_\text{s}$ = 1.32 Bohr with the molecular x = 0.5-0.7, both the pressure and conductivity go through abrupt changes, indicating a first-order phase transition. These results can be understood as follows. At the molecular fraction x = 0.8-0.85, the partially dissociated hydrogen atoms lead to a remarkable change of electronic structure of liquid hydrogen. However, these dissociated hydrogen atoms are surrounded mostly by freely rotating H$_2$ molecules, which raises the pressure upon dissociation \cite{tamblyn_prl_2010}. At higher pressures with a lower molecular fraction of x = 0.5-0.7, the H$_2$ molecules are no longer freely rotating and the dissociated hydrogen atoms can fill into the voids between the molecules, which lead to a drop in pressure upon dissociation \cite{tamblyn_prl_2010}. To summarize, the calculations of electric conductivity show that the liquid hydrogen has two separate electronic and atomic structural transition, which is qualitatively consistent with the above results that the metallization and dissociation of liquid hydrogen are not coinstantaneous.}

\begin{figure}
  \includegraphics[width=3.37 in]{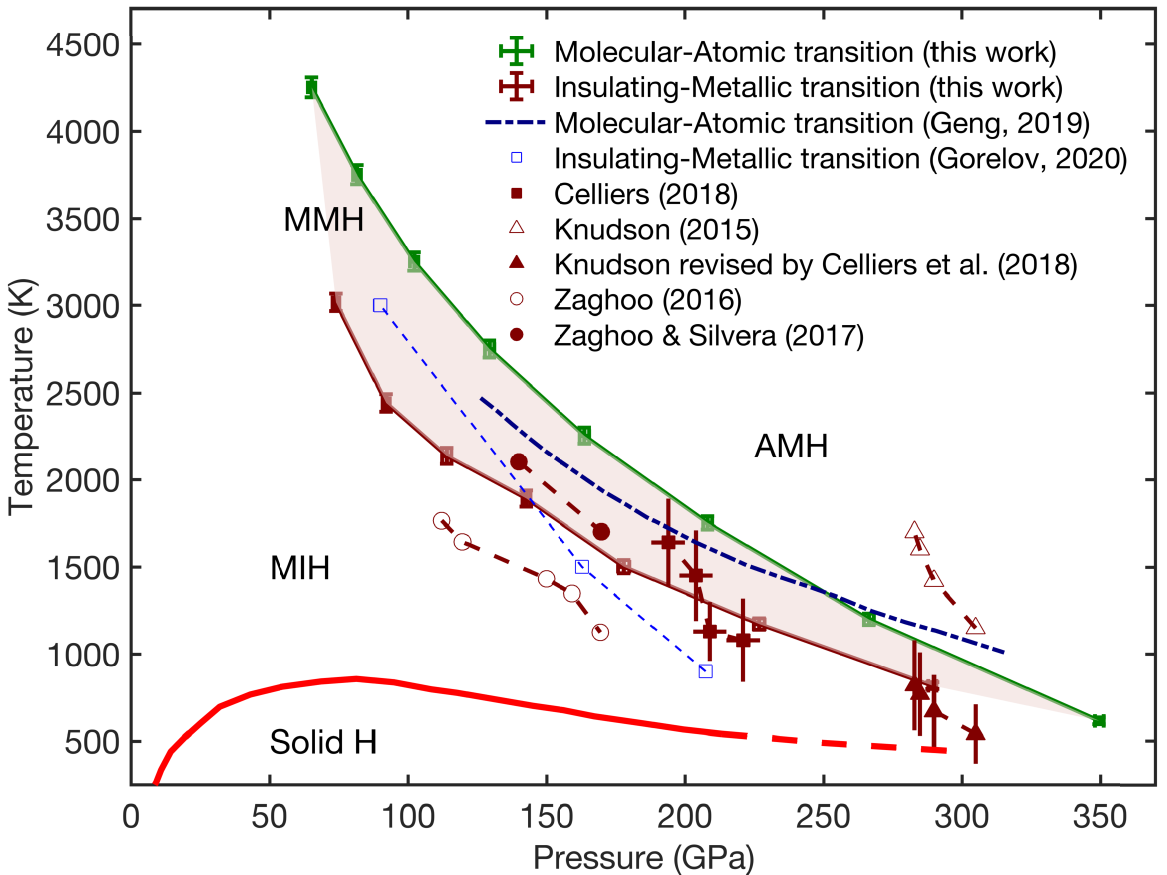}
  \caption{LLPT on the pressure-temperature phase diagram of hydrogen. Insulating-metallic transition and molecular-atomic transition boundaries determined in this work are shown as dark red and green solid lines, respectively. \textit{NVT} AIMD simulations of 512 hydrogen atoms were performed with the vdW-DF functional at different temperatures and densities from which the molecular fraction and pressure were calculated. The two transition boundaries were estimated by the temperature and the pressure that reproduce the molecular fraction x=0.8 (dark red line) and x=0.5 (green line), respectively. Futher details can be found in Fig. S6 of the Supplemental Material \cite{si}. MIH, MMH and AMH indicates the molcular-insulating hydrogen, molecular-metallic hydrogen and the atomic-metallic hydrogen in the liquid regime.  The light red line indicates the melting line adapted from Ref. \onlinecite{nellis_perspective_2021}. Symbols with dashed lines (dark red) show several experimental measurements of insulating-metallic phase transition. The dark blue dashed line shows the MAT phase boundary obtained by Geng et al. using the vdW-DF functional \cite{geng-prb-2019}. The light blue squares are IMT points calculated by Gorelov et al. \cite{gorelov-prb-2020} using quantum Monte Carlo. Recent experimental reports of the insulating-metallic transition boundary by Celliers et al. \cite{peter-science-2018}, Knudson et al. \cite{knudson-science-2015} and Zaghoo et al. \cite{zaghoo-prb-2016,zaghoo_pnas_2017} are presented by the dark red points.
}
  \label{fig:4} 
\end{figure}

To provide a more quantitative prediction, we draw the insulating-metallic ($x = 0.8$) and molecular-atomic ($x = 0.5$) phase transition boundaries in the pressure-temperature phase diagram, using AIMD simulations with the benchmarked vdW-DF functional (See Supplementary Material I and Fig. S6 for more details \cite{si}). 
We note that the two criteria may represent the lower/upper bound for IMT/MAT respectively on the transition pressure/temperature.  
However, it there two criteria are taken,
as shown by the pink shaded area in Fig. \ref{fig:4}, at a constant temperature the insulating-metallic transition pressure is shifted downward 30-60 GPa from the molecular-atomic transition pressure. 
At a constant pressure, the insulating-metallic transition temperature is lower than the molecular-atomic transition temperature by a range of 300 to 1000 K.
The two phase boundaries mean that liquid hydrogen can be divided into three phases, namely the molecular insulating hydrogen (MIH), the molecular metallic hydrogen (MMH), and the atomic metallic hydrogen (AMH), as labelled in Fig. \ref{fig:3} and Fig. \ref{fig:4}.
Molecular semi-metallic phase has been observed in the low temperature solid regime of dense hydrogen\cite{eremets_semimetallic_2019}, and partially dissociated molecular phases of solid hydrogen have also been discussed experimentally and theoretically \cite{dalladay-simpson_evidence_2016,monserrat_structure_2018}.
Here our prediction suggests that there is a wide range of molecular metallic hydrogen in the liquid regime, where we can further explore the intriguing electronic transitions of hydrogen.
We note that a different definition of molecular fraction was used in Ref. \onlinecite{geng-prb-2019} to determine the MAT boundary, and our estimates agree quite well with theirs, suggesting that the conclusions reached are insensitive to the definition of the molecular fraction.

In experiments of liquid hydrogen, most of the measures are monitoring the metallization instead of the structural phase transition.
Therefore, the separation of the two phase transition boundaries allows us to make further quantitative comparison with experimental data, some of which are also plotted in Fig. \ref{fig:4}.
Apart from the original results of Knudson et al., which was later re-interpreted by Celliers et al.\cite{peter-science-2018}, our new prediction of the insulating-metallic transition boundary are in better agreements with experimental data \cite{zaghoo_pnas_2017,peter-science-2018}.
Vibrational measurements using e.g. Raman and Infrared spectroscopy have been performed to identify structural transitions of dense solid hydrogen \cite{howie_raman_2015}, and further extension of such techniques are desired to determine the molecular-atomic LLPT.
On the contrary, theoretical studies of LLPT have mostly focused on structural phase transitions, and there is space for further developments of theoretical methods, e.g. employing other electronic structure calculations.
Nuclear quantum effects can also lead to a shift of both boundaries by a small amount \cite{morales-prl-2013,vandebund-prl-2021}, but it remains to be investigated whether nuclear quantum effects on MAT and IMT are effectively the same, and how sensitive are nuclear quantum effects to the choice of DFT functional.

To conclude, we have performed DMC calculations and AIMD simulations of liquid hydrogen at a wide range of pressures and temperatures.
Our main finding is that the LLPT of hydrogen is in fact two separate transitions that do not coincide with each other.
Although questions remain to be further investigated, such as (i) the nature of the two transitions, i.e. whether they are first-order or continuous phase transitions at thermodynamic limit; (ii) the exact transition pressures and temperatures on the phase diagram.  
Nevertheless, our study shows strong evidence that the insulting-metallic transition occurs before the molecular-atomic phase transition as the temperature and pressure increases, leading to a regime of molecular metallic hydrogen with a width of, at most, 30-60 GPa and 300-1000 K.
Our results provide an important addition to the current understanding of the phase diagram of hydrogen.
Last but not the least, the onset of metallic transition in molecular liquid hydrogen is encouraging for the experimental chasing of the holy grail of metallic hydrogen and the seeking of high temperature superconductor in dense hydrides.

~\\

\begin{acknowledgments}
The authors thank X.-Z. Li and A. Zen for helpful discussions.
This work was supported by the Strategic Priority Research Program of Chinese Academy of Sciences under Grant No. XDB33000000, and the National Natural Science Foundation of China under Grant No. 11974024, 92165101, 11935002. 
We are grateful for computational resources provided by TianHe-1A, the High Performance Computing Platform of Peking University, Shanghai Supercomputer Center, and the Platform for Data Driven Computational Materials Discovery of the Songshan Lake Materials Lab.
\end{acknowledgments}



\clearpage

\LARGE Supplementary Material
\normalsize

\makeatletter 
\renewcommand{\thefigure}{S\@arabic\c@figure}
\makeatother
\setcounter{figure}{0}

\title{Supplementary Material for "Onset of metallic transition in molecular liquid hydrogen"}

\date{\today}
\newcommand{\bc}[1]{{\color{blue} #1 }}


\section{I. Computational details}
\subsection{\textit{Ab initio} molecular dynamics simulation}
Density functional theory (DFT) based \textit{ab initio} molecular dynamics (AIMD) simulations were carried out using the Quantum Espresso (QE) package \cite{giannozzi-jpcm-2009,giannozzi-jpcm-2017} with PBE \cite{perdew-prl-1996}, PBE+vdW \cite{tkatchenko-prl-2009}, PBE0 \cite{adamo_pbe0_1999}, BLYP-D2 \cite{grimme_blypd2_2006}, vdW-DF \cite{dion_vdwdf_2004}, vdW-DF2 \cite{klime_vdwdf2_2009}, SCAN \cite{sun-prl-2015} and SCAN+rvv10 \cite{vydrov_jcp_2010,sabatini_prb_2013} functionals.
The interactions between the valence electrons were treated with Hamann-Schl\"uter-Chiang-Vanderbilt pseudopotentials \cite{hamann-prl-1979,vanderbilt-prb-1985}. Each simulation was performed in the canonical ($NVT$) ensemble, and the temperature was controlled by a Stochastic Velocity Rescaling thermostat \cite{bussi2007svr}.
A time step of 0.24 fs was used for the molecular dynamics.
A cutoff energy of 80 Ry was used for the plane-wave expansions.
All the simulations were performed in a cubic simulation cell of 128 and 512 H atoms with ($3 \times 3 \times 3$) Monkhorst-Pack k-points.
\subsection{Estimate of transition boundaries on the P-T phase diagram}
To estimate the molecular-atomic transition and the insulating-metallic transition boundaries given by the molecular fraction x=0.5 and x=0.8, we carried out systematic AIMD simulations using the benchmarked vdW-DF functional. The simulations were performed in the NVT ensemble with 512 hydrogen atoms in a cubic box. The density and temperature are varied and short simulations were performed to obtain an estimate of molecular fraction x. If the molecular fraction x is within a difference threshold of 0.02 to the targeted x=0.5 and x=0.8 values, then the simulations were continued for ~1 ps to calculate the pressure and molecular fraction. The data are presented in Fig. S6.
\subsection{Density functional theory calculation}
Additional DFT total energy calculations for benchmark with PBE, PBE+vdW, PBE0, vdW-DF, vdw-DF2 and BLYP-D2 functionals were performed with the QE package. A cutoff energy of 300 Ry was used for the plane-wave expansions. DFT calculations with SCAN, SCAN+rvv10 the HSE06 functionals were performed with the VASP code \cite{kresse_vasp_1996}. Projector augmented wave (PAW) potentials along with a 500 eV cutoff energy was employed for the expansion of the electronic wavefunctions. All the DFT calculations used a ($3 \times 3 \times 3$) Monkhorst-Pack grid with system sizes of 128 and 512 hydrogen atoms. Larger k-points up to ($12 \times 12 \times 12$) were tested using DFT-PBE, and no significant differences in the results of our DFT calculations were found.
\subsection{Diffusion Monte Carlo calculation}
DMC calculations were performed using the CASINO package \cite{needs-jpcm-2009}, with the size-consistent DMC algorithm ZSGMA \cite{zen-prb-2016}. The recently developed energy consistent correlated electron pseudopotentials (eCEPPs) \cite{trail-jcp-2017} were used. The trial wave functions were of the Slater-Jastrow type where the Jastrow factor contains the electron-nucleus, electron-electron, and electron-electron-nucleus terms. The Slater single particle wave functions were obtained from the local-density approximation (LDA) plane-wave DFT calculations with a plane-wave cutoff of 300 Ry. using the QE package and then expanded in terms of B-splines \cite{alfe-prb-2004}. 
A time step of $\tau = 0.01$ a.u. and twist number of 24 were used in our DMC calculations.
The above setup was tested to be essentially converged in calculating the relative energies of liquid hydrogen system in previous work \cite{chen_jcp_2014}.

\subsection{Metadynamics with machine learning force field}

To efficiently generate configurations with large structural variety,
we employed well-tempered metadynamics simulations~\cite{barducci2008well},  using the machine learning potential provided by Cheng et al. \cite{cheng-nature-2020}.
The collective variable for driving the biased simulation is the number of molecular hydrogen atoms, which is the ``cn.between-1'' order parameter detailed in the next section.
We performed NVT simulations with the system size of 128 H atoms, at five densities ($\text{r}_\text{s}$= 1.30, 1.35, 1.40, 1.45, 1.50 Bohr), and at each density five temperatures (T= 1200, 1500, 1800, 2000, 2500 K) were considered.
The setup for the metadynamics bias is the same as Ref.~\cite{cheng-nature-2020}.

\subsection{Electrical conductivity calculation}

The frequency dependent dielectric function $\epsilon(\omega)$ of liquid hydrogen is given by DFT calculations via QE package. The dc electric conductivity $\sigma$ is then calculated at the zero frequency limit by the Kubo–Greenwood equation\cite{kubo1957sigma,gajdo2006sigma},
\begin{equation}
    \sigma(\omega) = \omega\times\frac{\mathrm{Im} \epsilon(\omega)}{4\pi}
\label{eq:llpt_epsi}
\end{equation}
By randomly taking 100 snapshots (to ensure a well-sampling of the electric properties) from each AIMD simulation, we calculate the evolution of $\sigma$ as a function of density, as shown in Fig. 4 of the main text and Fig. \ref{fig:eos800},\ref{fig:eos1000_128H}.

\section{II. Definition of molecular fraction \label{sec:op}}
\begin{figure}[htbp]
  \includegraphics[width=4.69 in]{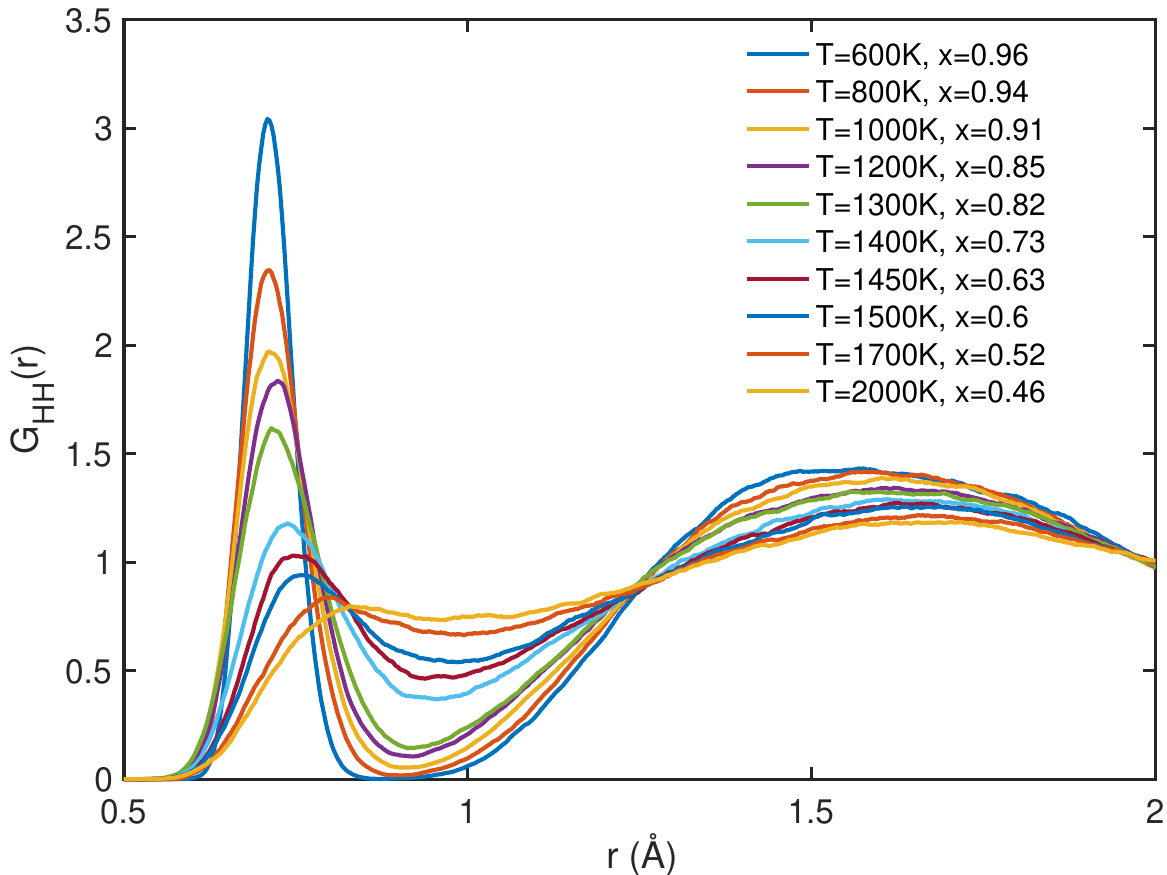}
  \caption{HH pair correlation functions $\text{G}_\text{HH}(\text{r})$ at different temperatures using vdw-DF AIMD simulations at the density of $\text{r}_\text{s}$ = 1.40 Bohr. The average molecular fraction x of liquid hydrogen in each simulation is also calculated and presented. The molecular-atomic phase transition (MAT) can also be observed by the vanishing of first peak of $\text{G}_\text{HH}(\text{r})$. We find that the vanishing of the first peak happens around x = 0.5 at different densities (including Fig. 1(d) in the main text), which is consistent with our MAT criterion in the main text.}
  \label{fig:s1} 
\end{figure}

To calculate the molecular fraction, we use the definition reported in Ref. \onlinecite{cheng-nature-2020}. 
The fraction of molecular H atoms is defined as the fraction of atoms with one neighbor within a smooth cutoff function that starts at 0.8~\AA{} and decays to zero at 1.1~\AA{}.
The cutoff function was selected based on the location of the first peak at roughly 0.75~\AA{}, and the first neighbor shell radius of the HH pair correlation functions at about 1~\AA{} (see Fig.~\ref{fig:s1}), such that the bonded hydrogen atoms within the first shell are counted.
We then select the atoms that have about one bonded neighbor, again using a smooth switching function.
The number of the molecular hydrogen atoms is implemented and calculated using PLUMED \cite{tribello_plumed_2014},
with the specification as the following:
\begin{verbatim*}
COORDINATIONNUMBER ...
LABEL=cn
SPECIES=1-128
SWITCH={CUBIC D_0=0.8 D_MAX=1.1}
BETWEEN1={GAUSSIAN UPPER=1.2 LOWER=0.8 SMEAR=0.2}
... COORDINATIONNUMBER
\end{verbatim*}
The number of the molecular hydrogen can then be fetched using the label ``cn.between-1'',
which sums up the molecular order parameter of each atom in the system,
and the molecular fraction is this number divided by the total H atoms.

\section{III. Clustering based agnostic identification of molecular and atomic hydrogen}

To further validate that the molecular fraction described in Sec.~II can be used as a reasonable order parameter to describe the molecular-atomic transition of liquid hydrogen, we carry out clustering based agnostic classification analysis. 

The principal component analysis (PCA) maps in Figure~\ref{fig:pca}~(a-c)
show 2D projections of the local environments of each H atom in dense hydrogen, based on the atomic Smooth Overlap of Atomic Positions (SOAP) descriptors~\cite{bart+13prb}.
We set the Gaussian width to be 0.2\AA, the radius of the atomic environment to be $r_\text{c}=1.2$ \AA, so that it includes the first neighbor shell, and expand the SOAP descriptor up to $l_\text{max}=8$ and $n_\text{max}=6$. 
We have tested other SOAP hyper-parameters, up to $r_\text{c}=2.2$ \AA, and the results are similar.
The snapshots of the hydrogen system were selected from the metadynamics simulations with 128 H atoms at rs=1.5 and T=1200K, 1800K and 2500K.
In Figure~\ref{fig:pca}~(a) we color each dot on the PCA map, which indicate one atom, by the molecular order parameter defined in the previous section.
(b) is similar to (a), except that we set a rigid cutoff of 0.5 for classifying the "molecular" and "atomic" hydrogen atoms.

In (c), we show an agnostic classification using the k-means clustering algorithm, based on the first 10 PCA components of the SOAP descriptors. 
The number of the clusters was hard-coded to be two.
By comparing (b) and (c), it can be seen that the agnostic classification produced very similar results as our ``hand-crafted'' molecular order parameter.
More quantitatively, (d) shows the frequency counts of the order parameters of the H atoms in each of the clusters:
the atoms that belong to the cluster 0 have order parameters that sharply peak at 1 (molecular), and the atoms in the cluster 1 predominantly have an order parameter of 0 (atomic).

Based on these analyses, we conclude that the molecular fraction can be used as a good order parameter and the 0.5 is a good criterion to separate the molecular and the atomic phase of liquid hydrogen.

\section{IV. Benchmark structure data set}
The benchmark liquid hydrogen structures are collected from two sets of simulations.
The first set is from the aforementioned metadynamics simulations, and
8 structures were selected from the simulations at 1200-2500K and $\text{r}_\text{s}$= 1.50 Bohr.
The second set is from
AIMD simulations using vdW-DF functional with the system size of 128 H atoms,
at the simulations at 1200-3000K and $\text{r}_\text{s}$= 1.50 Bohr (see Fig. 1(d) of the main text).
6 structures were selected from this set.
In total, 14 benchmark structures were then chosen from these trajectories with a molecular fraction between 0.3 and 1.0 in a step of $\sim$ 0.05.
The structure files are available online.

\clearpage

\section{V. Supplementary figures}

\begin{figure}
  \includegraphics[width=6.69 in]{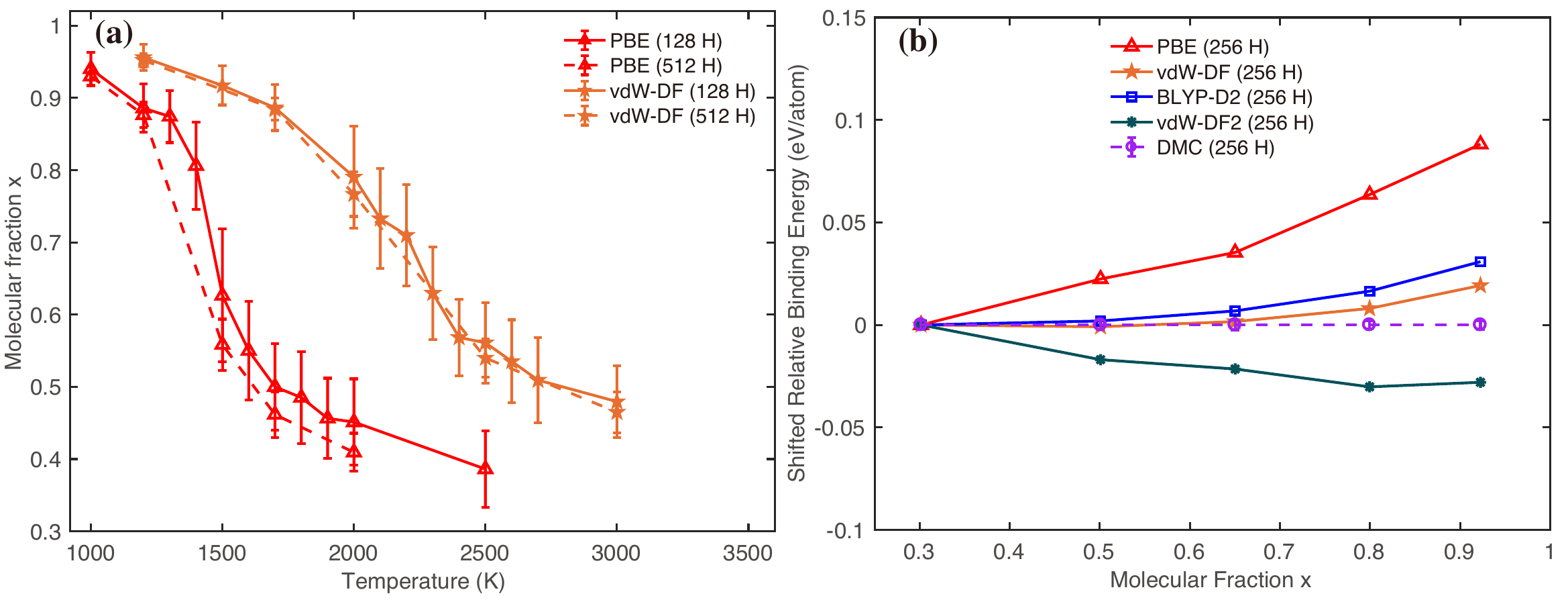}
  \caption{Tests on the size effects. (a) Molecular fraction x as a function of temperature calculated from AIMD simulations of 128 and 512 hydrogen atoms. No statistical differences were found at the larger system size, which indicates the size effect can be neglected in our simulations. (b) Shifted relative binding energies as a function of molecular fraction calculated by different functionals with a larger size of 256 hydrogen atoms. The structures are collected from vdW-DF AIMD simulations of 256 hydrogen atoms with a density of rs=1.5 Bohr at different temperatures. The shifted relative binding energies with respect to DMC results are aligned by subtracting the value at the atomic side (x=0.3). Again, the results are consistent with Fig. 2 of the main text, which indicates a minor size effect.
 }
  \label{fig:s2} 
\end{figure}

\begin{figure}
  \includegraphics[width=6.69 in]{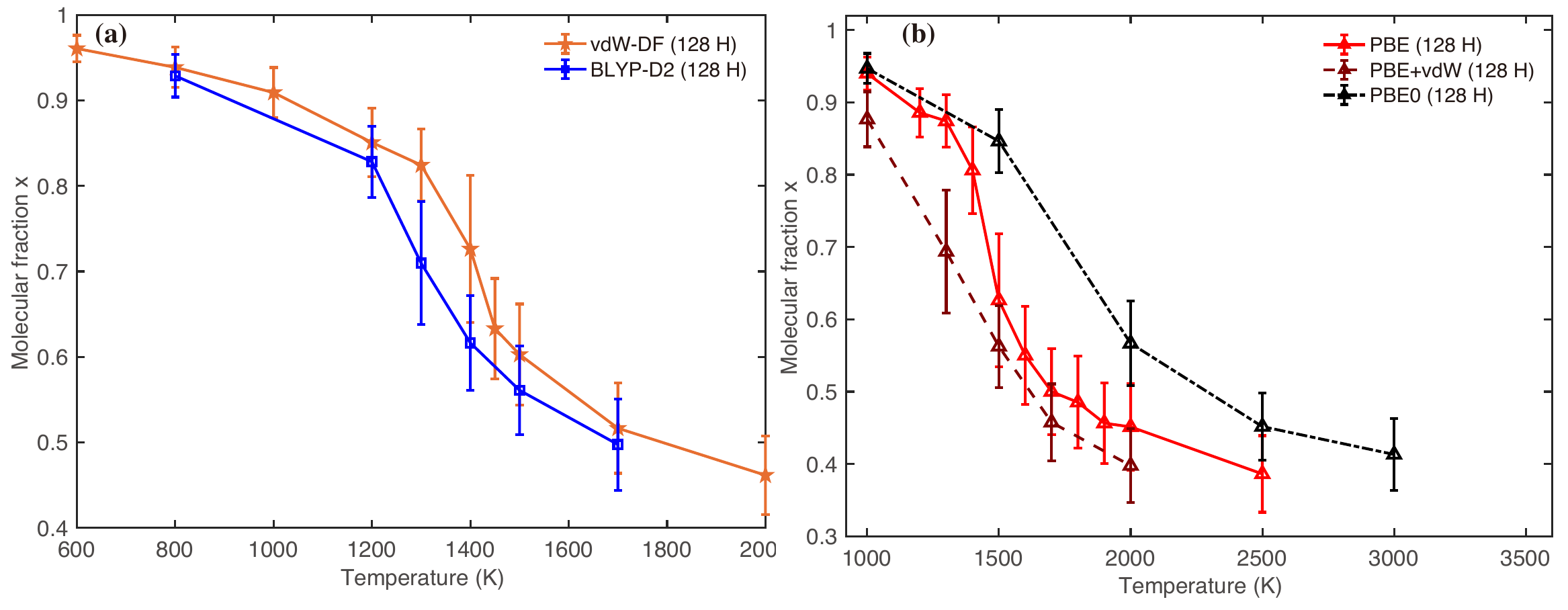}
  \caption{Molecular fraction x calculated by different functionals. (a) Comparison of vdW-DF to BLYP-D2 at the density of $\text{r}_\text{s}$ = 1.40 Bohr. The two results are close to each other, which is similar to the behavior in Fig. 1 (c) of the main text. The consistency of these two functionals verifies the accuracy of our DMC benchmark. (b) Results calculated using PBE, PBE+vdW and PBE0. Van der Waals correction has a minor influence in describing the molecular-atomic transition, while the choice of underlying exchange correlation functional is dominant.
 }
  \label{fig:s3} 
\end{figure}

\begin{figure}
  \includegraphics[width=6.69 in]{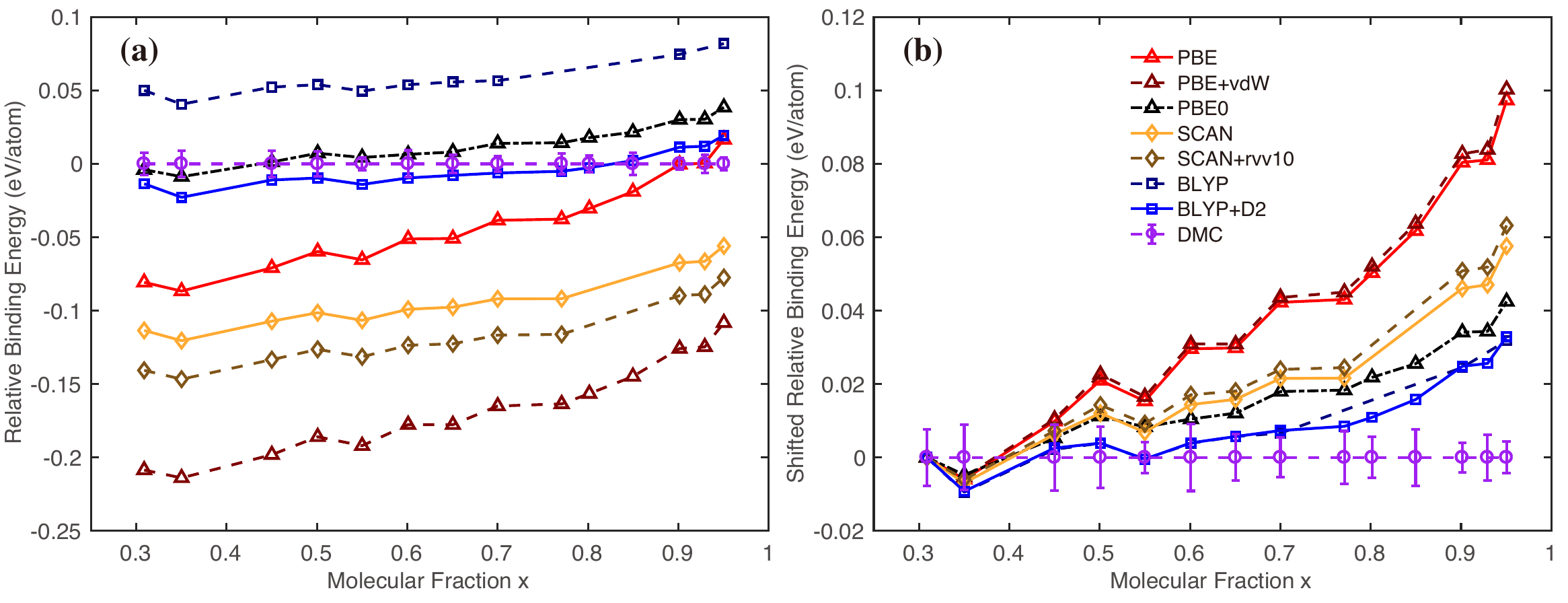}
  \caption{Tests on the van der Waals correction. (a) Relative binding energies as a function of molecular fraction calculated by different functionals, including different van der Waals corrections. (b) The shifted relative binding energies with respect to DMC results are aligned by subtracting the value at the atomic side (x=0.3). Again, van der Waals correction has a minor influence in describing the relative binding energies at different molecular fractions, while the choice of underlying exchange correlation functional is dominant. 
 }
  \label{fig:s4} 
\end{figure}

\begin{figure}
  \includegraphics[width=6.69 in]{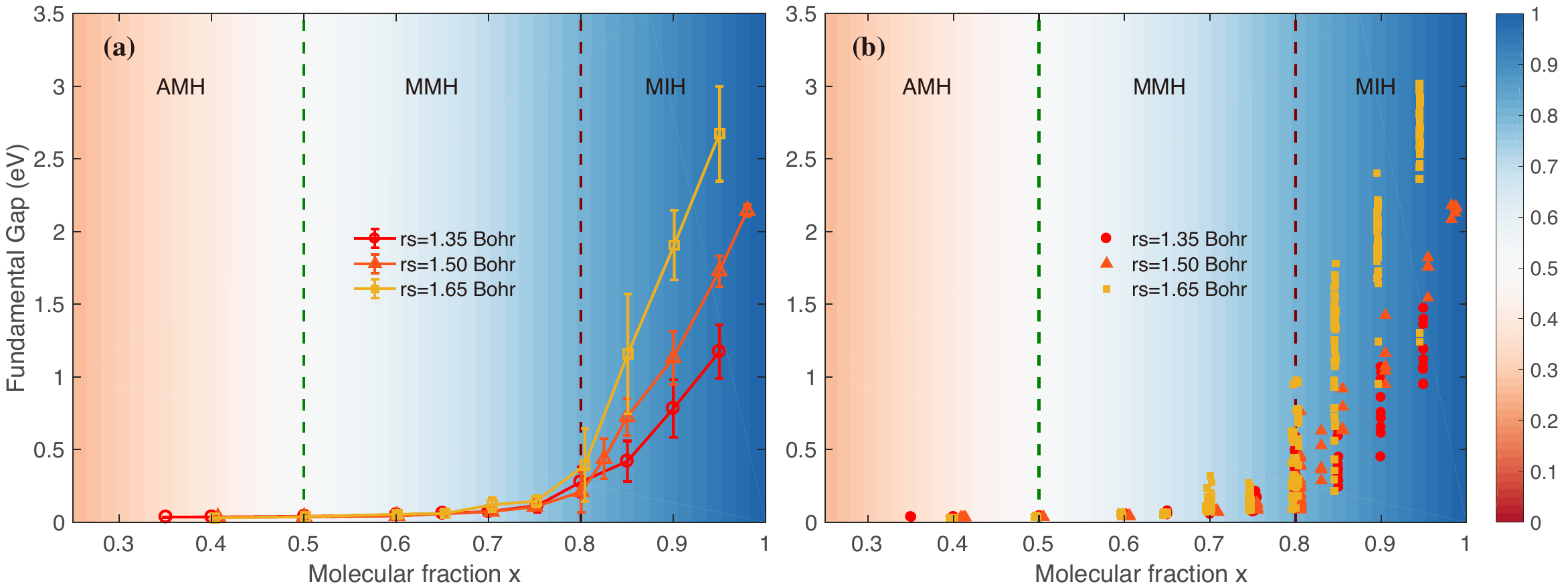}
  \caption{(a) Fundamental energy gap for liquid hydrogen as a function of molecular fraction x calculated by DFT-PBE methods with 512 hydrogen atoms at different densities. (b) Scatter plot of data points used in (a). The insulating-metallic phase transition boundary at molecular fraction x=0.8 is robust for all the different densities.}
  \label{fig:s5} 
\end{figure}

\begin{figure}
  \includegraphics[width=4.69 in]{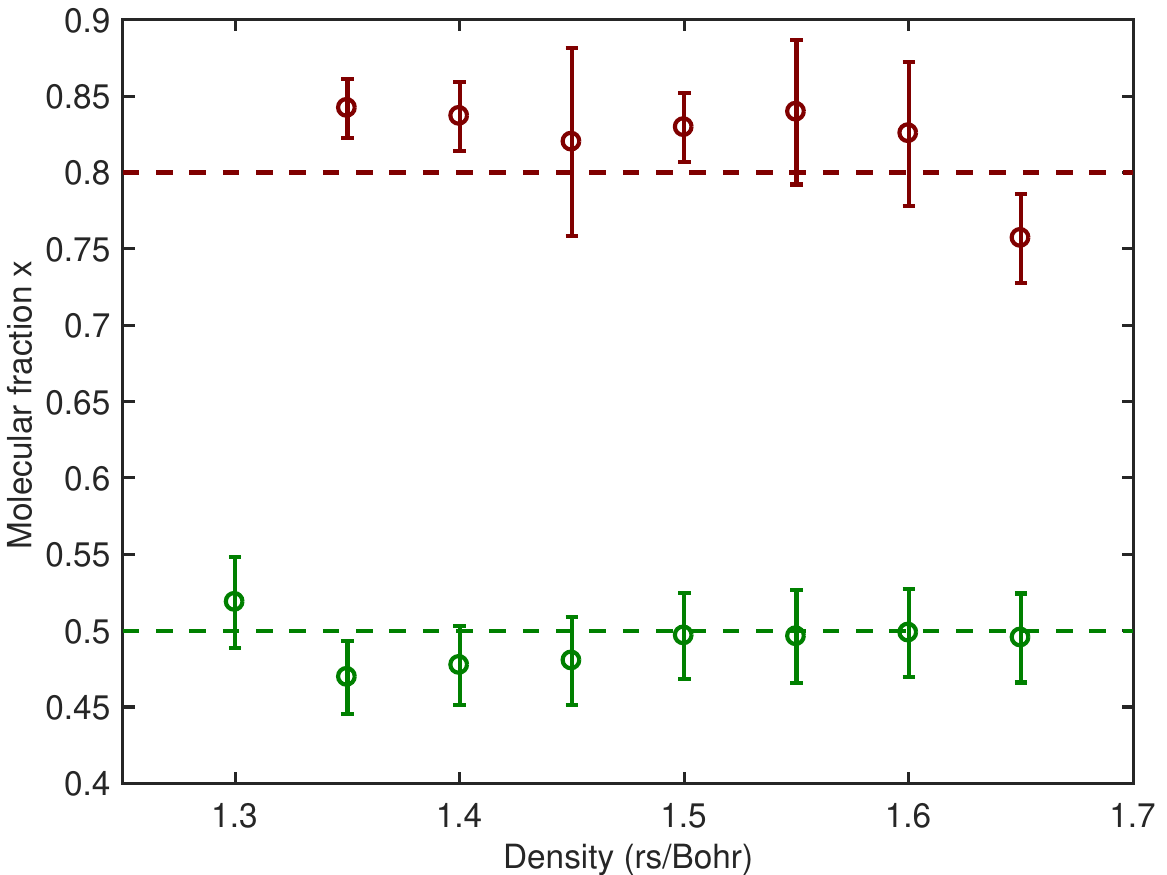}
  \caption{The corresponding molecular fraction of the data points used in Fig. 4 of the main text. AIMD simulations were performed in NVT ensemble, where the pressure and the molecular fraction were calculated after reaching equilibrium in each simulation. The green points describe the insulating-metallic transition boundary with a molecular fraction x around 0.5, and the corresponding temperature and pressure from left to right are (615 K; 349.48 GPa), (1200 K; 266.51 GPa), (1750 K; 208.18 GPa), (2250 K;163.75 GPa), (2750 K; 129.34 GPa), (3250 K; 102.25 GPa), (3750 K; 81.70 GPa), and (4250 K; 65.35 GPa). The dark red points describe the molecular-atomic transition boundary with a molecular fraction x around 0.8, where the corresponding temperature and pressure from left to right are (820 K; 289.40 GPa), (1170 K; 226.66 GPa), (1500 K; 178.04 GPa), (1890 K; 142.65 GPa), (2140 K; 114.03 GPa), (2440 K; 92.14 GPa), and (3020 K; 74.13 GPa).}
  \label{fig:s6} 
\end{figure}

\begin{figure*}
\includegraphics[width=6.69 in]{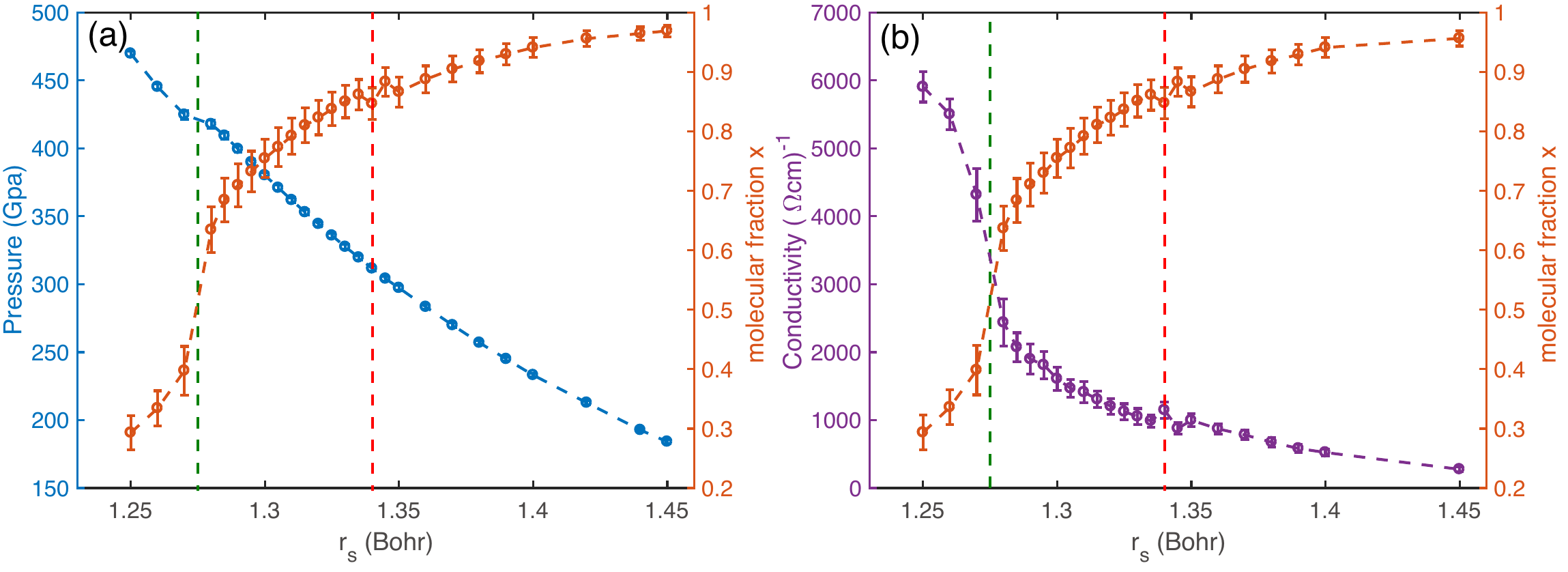}
\caption{
 {Pressure and conductivity of liquid hydrogen from AIMD simulations. All the simulations were carried out at the density of $\text{r}_\text{s}$ = 1.30-1.45 Bohr in the \textit{NVT} ensemble with 256 hydrogen atoms at the temperature of 800 K.
 The calculated pressure (a) and conductivity (b) as a function of density. The corresponding molecular fractions are also plotted. The vertical green and red dashed lines represents the sudden change of pressure and conductivity respectively.}
 }
\label{fig:eos800} 
\end{figure*}

\begin{figure*}
\includegraphics[width=6.69 in]{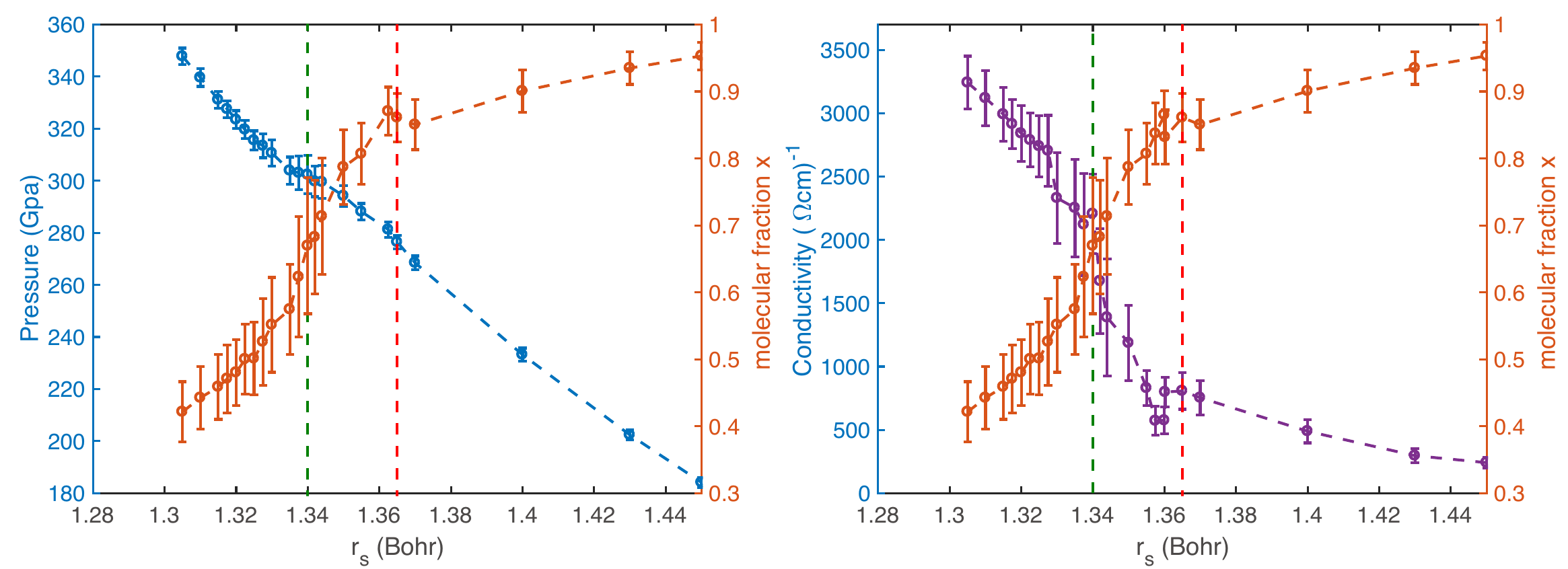}
\caption{
 {Pressure and conductivity of liquid hydrogen from AIMD simulations. All the simulations were carried out at the density of $\text{r}_\text{s}$ = 1.30-1.45 Bohr in the \textit{NVT} ensemble with 128 hydrogen atoms at the temperature of 1000 K.
 The calculated pressure (a) and conductivity (b) as a function of density. The corresponding molecular fractions are also plotted. The vertical green and red dashed lines represents the sudden change of pressure and conductivity respectively.}
 }
\label{fig:eos1000_128H} 
\end{figure*}

\begin{figure}
\includegraphics[width=6.69 in]{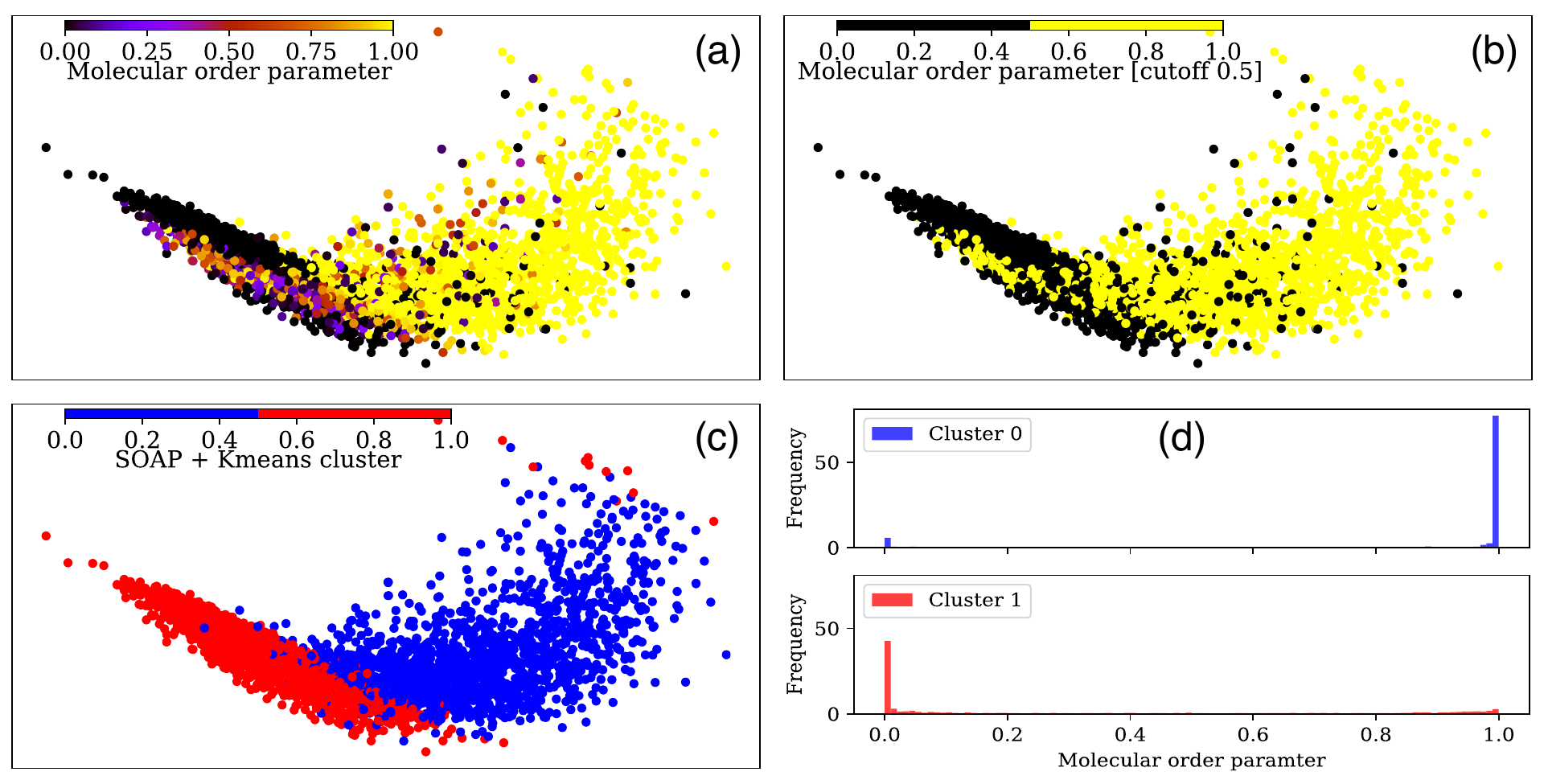}
  \caption{Comparison between the order parameter used to determine the molecular fraction, and clustering results based on SOAP descriptors.
  (a) is the principal component analysis (PCA) map, colored using the molecular order parameter
  (b) is same as (a) but used a discrete color scale with a cutoff of 0.5.
  (c) is the PCA map colored according to the k-means clustering based on the SOAP descriptors.
  (d) shows the distributions of the order parameters of the atoms that belong to each of the clusters identified in (c).
 }
  \label{fig:pca} 
\end{figure}

\clearpage

\bibliography{ref}
\end{document}